\documentclass[prl,aps,twocolumn,showpacs]{revtex4}

\usepackage{pdfsync}
\usepackage[english]{babel}
\usepackage{latexsym}

\usepackage{amsmath}

\usepackage{amsfonts}

\usepackage{amssymb}
\usepackage{amscd}
\usepackage{amsgen,amstext,amsbsy,amsopn}
\usepackage{math rsfs}

\newcommand{\bdm}{\begin{displaymath}}
\newcommand{\edm}{\end{displaymath}}
\newcommand{\bdn}{\begin{eqnarray}}
\newcommand{\edn}{\end{eqnarray}}
\newcommand{\bay}{\begin{array}{c}}
\newcommand{\eay}{\end{array}}
\newcommand{\ben}{\begin{enumerate}}
\newcommand{\een}{\end{enumerate}}
\newcommand{\beq}{\begin{equation}}
\newcommand{\eeq}{\end{equation}}
\newcommand{\beqq}{\begin{equation*}}
\newcommand{\eeqq}{\end{equation*}}

\newcommand{\lf}{\left}
\newcommand{\ri}{\right}

\newcommand{\xv}{\mathbf{x}}

\newcommand{\rv}{\mathbf{r}}

\newcommand{\diff}{\mathrm d}
\newcommand{\eps}{\varepsilon}

\newcommand{\gpf}{\mathcal{E}^{\mathrm{GP}}}

\newcommand{\magnp}{\mathbf{A}}

\newcommand{\half}{\hbox{$\frac12$}}

\newtheorem{teo}{Theorem}
\begin{document}

\title{Rotating superfluids in anharmonic traps: \\ From vortex lattices to giant vortices}

\author{Michele Correggi}
\affiliation{Dipartimento di Matematica,
Universit\`{a} degli Studi Roma Tre,
Largo San Leonardo Murialdo 1, 00146, Roma, Italy.}
\author{Florian Pinsker}
\affiliation{DAMTP, University of Cambridge, Wilbertforce Road, Cambridge CB3 0WA, United Kingdom.}
\author{Nicolas Rougerie}
\affiliation{CNRS and Universit\'e de Cergy-Pontoise, D\'{e}partement de Math\'{e}matiques, CNRS-UMR 8088, Site de Saint Martin, 2 avenue Adolphe Chauvin, 95302 Cergy-Pontoise Cedex,
France.}
\author{Jakob Yngvason}
\affiliation{Faculty of Physics,
University of Vienna, Boltzmanngasse 5
}
\affiliation{Erwin Schr{\"o}dinger Institute for Mathematical Physics, Boltzmanngasse 9, A-1090 Vienna, Austria.}

\date{October 6, 2011}

\begin{abstract}We study a superfluid in a rotating anharmonic trap and explicate a rigorous proof of a transition from a vortex lattice to a giant vortex state as the rotation is increased beyond a limiting speed determined by the interaction strength. The transition is characterized by the disappearance of the vortices from the annulus where the bulk of the superfluid is concentrated due to centrifugal forces while  a macroscopic phase circulation remains. 
The analysis is carried out within two-dimensional Gross-Pitaevskii theory at large coupling constant 
and reveals significant differences between \lq soft' anharmonic traps (like a quartic plus quadratic trapping potential) and traps with a fixed boundary: In the latter case the transition takes place in a parameter regime where the size of vortices is very small relative to the width of the annulus whereas in \lq soft' traps the vortex lattice persists until the width of the annulus becomes comparable to the vortex cores. Moreover, the density profile in the annulus where the bulk is concentrated 
is, in  the \lq soft'  case, approximately gaussian with long tails and not of the Thomas-Fermi type like in a trap with a fixed boundary. 
\end{abstract}
\pacs{03.75.Hh, 47.32.-y, 47.37.+q}

\maketitle
\section{I. Introduction}
A superfluid confined in a rotating trap undergoes several phase transitions as the rotational speed is increased. In anharmonic traps, where the speed can in principle be arbitrarily large, these transitions can essentially be associated with   {\it three critical speeds}.  At slow rotation the fluid is vortex free \cite{IM1,AJR} but when the speed exceeds a first critical value a quantized vortex is created.  As the speed goes further up the number of vortices increases \cite{A, Fe1,Co, CD, AD, IM2}  and a vortex lattice emerges\footnote{The emergence of a vortex lattice implies, among other things,  the solutions considered in \cite{BB1, BB2, BBS} can not be ground states of the interacting system at rapid rotation. We thank the referee for pointing out these references to us.}.This picture holds in harmonic and anharmonic traps alike, but in the latter case a new feature comes into play when a second critical velocity is exceeded: The centrifugal forces create a \lq hole' with strongly depleted density around the center of the trap \cite{FB, CDY1, CY} while the vortex lattice still prevails in the bulk.  At a third critical speed a remarkable transition takes place: The vortex lattice disappears and the fluid becomes again vortex free in the bulk. Thus the process described above is in a sense reversed. All vorticity is now concentrated in a {\em giant vortex} situated in the hole and creating a macroscopic phase circulation in the bulk. 
In the past several authors have studied this phenomenon theoretically by variational and numerical methods \cite{Fe2, FJS,FB,FZ,KTU,KF, DK,KB} but mathematically rigorous proofs of the giant vortex transition have been obtained only very recently \cite{R1, CRY, CPRY1, CPRY2, R2}. An experimental realization of this transition appears to be still out of reach although anharmonic traps have been available already for some time \cite{BSSD,Ry,H,Sh}.

In this paper we present rigorous results 
on the giant vortex transition in a two-dimensional trapping potential that is the sum of a quadratic and a homogeneous potential of the form
\beq\label{pot} V_{\rm trap}(r)=kr^s+\half\Omega_{\rm osc}^2r^2\eeq
with $r$ the radial variable, $k>0$, $s>2$ and
 $0\leq \Omega_{\rm osc}<\Omega_{\rm rot}$ where $\Omega_{\rm rot}$ denotes the rotational speed.
The case $s=4$ was studied in \cite{R1} in an asymptotic regime that corresponds to a {\em fixed} value of the interaction strength while the rotational speed tends to infinity. 
In contrast, the papers \cite{CRY,CPRY1, CPRY2, R2} focus on the interplay between rotation speed and interaction strength and provide precise information about the third critical speed as a function of the interaction parameter when the latter is large. The model studied in these papers is that of a \lq flat' trap with the unit circle as boundary as in \cite{CDY1, CY} that can formally be regarded as the limiting case $s=\infty$.  A mathematical advantage of this model is that the extension of the system is fixed, while for finite $s$ the system expands as $\Omega$ and/or $\eps^{-1}$ tend to $\infty$. There are, however, both physical and mathematical reasons for treating the latter case separately. One reason is that $s=4$ corresponds to the lowest correction beyond quadratic in the Taylor expansion of a symmetric potential around its minimum and such a deviation from a quadratic potential has a better chance to be realized in experiments than the limiting case $s\to\infty$.  A further reason is that the limit $s\to\infty$ can {\em not} be interchanged with the limit of strong coupling which is the basis of almost all rigorous analysis of quantized vortices including the present one. This will be explained further below. In particular, the formulas for the limiting velocities obtained in this paper do not simply  pass in the $s\to\infty$ limit over to those for a flat trap with Dirichlet boundary conditions as considered in \cite{CPRY1}, contrary to what might be expected. 

In fact, our analysis reveals significant qualitative differences between the two cases. In the flat trap the giant vortex transition takes place at a rotational velocity where the vortex cores are still vanishingly small relative to the width of the annulus containing the bulk of the density. In a trap with finite $s$, on the other hand, the vortex lattice persists until the width of the annulus becomes comparable to the size of the vortex cores. Moreover, the density profile in the flat trap is well approximated by a \lq Thomas-Fermi' (TF) functional without a kinetic energy term, while in the case of finite $s$ the radial kinetic energy can not be neglected and the profile in the radial variable is approximately  gaussian centered at the middle of the annulus. The long tails of such a function impose the use of a larger domain than in former situations \cite{AAB,CRY,CPRY1}. All together these differences necessitate new ideas for the proofs of the giant vortex transition compared to the earlier papers. 
Common to the present setting and \cite{CRY,CPRY1} is a macroscopic phase circulation around the annulus as well as breaking of rotational symmetry of the density in the ground state even in the giant vortex regime.

The mathematical proofs of some of the statements in the sequel are rather lengthy and will not be detailed in the present paper that is concerned with essential ideas and the main results. A full account of the proofs can be found in  \cite{CPRY2}.

\section{II. The mathematical setting}
We now define precisely the mathematical setting which is that of two-dimensional Gross-Pitaevskii (GP) theory, cf.\ \cite{A}. The general form of the energy functional for the superfluid order parameter $\Psi$ (wave function of the condensate)  in a rotating trap is
 \begin{multline}
	\label{GPfunctional}
	\gpf_{\rm phys}[\Psi] = \int_{\mathbb R^2} {\diff} \rv \: \lf\{ \half\lf|\lf( {\rm i} \nabla +\magnp \ri) \Psi \ri|^2 \right.\\ \left.+(V_{\rm trap}- \half \Omega_{\rm rot}^2 r^2) |\Psi|^2 + \frac{|\Psi|^4}{\eps^2} \ri\}.
\end{multline}
Here  $\magnp=\Omega_{\rm rot}\mathbf{e}_3 \wedge \rv$ with $\Omega_{\rm rot}>0$  the rotation velocity, $ \mathbf e_3$ the unit vector in the $x_3$-direction, $\rv=(x_1,x_2)$, $r=|\rv|$,  $V_{\rm trap}$ the trap potential and  $1/\eps^2$  with $\eps>0$ the GP coupling constant. The latter is, for a dilute Bose gas,  given by $2\pi Na/L$ with $N$ the particle number, $a$ the scattering length of the interaction potential between the particles and $L$ a characteristic length in the $x_3$-direction \cite{LS}. Units have been chosen such that $\hbar$ and the particle mass are both 1. (This differs from \cite{CRY, CPRY1} where the particle mass is taken to be $\half$). The normalization of the wave function is $\int|\Psi|^2=1$. The subscript \lq\lq phys" indicates that the functional \eqref{GPfunctional} is written in the terms of the original physical variables, in contrast to the scaled functional defined in \eqref{gpscaled} below.
We denote by $E^{\rm GP}_{\rm phys}$ the GP energy, i.e., the minimum of \eqref{GPfunctional} under the normalization condition, and by $\Psi^{\rm GP}$ any of the (in general non-unique) minimizers. In this paper we always assume {\em strong coupling}, which means $\eps\ll1$.

We now specialize to external potentials of the form \eqref{pot} and  $\Omega_{\rm rot}>\Omega_{\rm osc}\geq 0$.
 With the definition  $\Omega_{\rm eff}=(\Omega_{\rm rot}^2-\Omega_{\rm osc}^2)^{1/2}$ the sum of the external and centrifugal potentials in \eqref{GPfunctional} becomes
 \beq\label{effpot1} V_{\rm trap}(r)-\half \Omega_{\rm rot}^2 r^2=kr^s- \half \Omega_{\rm eff}^2 r^2.\eeq
The limiting case $s\to\infty$ and $ \Omega_{\rm osc}=0$ corresponds to the \lq flat' trap considered in \cite{CDY1, CY, CRY, CPRY1}; the potential \eqref{effpot1} is then simply $-\half\Omega_{\rm rot}^2r^2$ and the integration is limited to the unit disc in $\mathbb R^2$.

In order that the effect of the quadratic term in the potential is visible also when $\Omega_{\rm rot}\to\infty$ it is natural to keep the ratio 
$\Omega_{\rm eff}/\Omega_{\rm rot}$ fixed and we  write accordingly \beq\Omega_{\rm eff}^2=\gamma\Omega_{\rm rot}^2\eeq with $0<\gamma\leq 1$ fixed. The potential $(kr^s-\half \gamma\Omega_{\rm rot}^2r^2)$ has a unique minimum at $r=R_{\rm m}$ with
\beq R_{\rm m}\label{Rm}=\left(\frac{\gamma\Omega_{\rm rot}^2}{sk}\right)^{1/(s-2)}.\eeq
We now write 
\beq\label{scaling}\rv=R_{\rm m}\mathbf x, \,\, r=R_{\rm m}x,\,\, \Psi(\mathbf r)=R_{\rm m}^{-1}\psi(\mathbf x),\,\, \Omega_{\rm rot}=R_{\rm m}^{-2}\Omega\eeq and obtain
\beq\mathcal E^{\rm GP}_{\rm phys}[\Psi]=R_{\rm m}^{-2}\mathcal E^{\rm GP}[\psi]\eeq
with the scaled energy functional
\begin{multline}\label{gpscaled}\mathcal E^{\rm GP}[\psi]=\int_{\mathbb R^2}\left\{\half|(\mathrm i\nabla+\mathrm \Omega x\mathbf e_{\vartheta})\psi|^2\right.\\ \left. +\gamma\Omega^2V(x)|\psi|^2+\eps^{-2}|\psi|^4\right\}\diff^2\mathbf x\end{multline}
where we have written \beq\label{effpot2} V(x)=\left(\hbox{$\frac1{s}$}x^s-\half x^2\right).\eeq
Note that the scaled potential $\gamma\Omega^2V(x)$ has a unique minimum at $x=1$, independently of $\Omega$, while the minimum of  \eqref{effpot1} wanders to infinity as $\Omega_{\rm rot}\to\infty$. Note also that we can take $\Omega\to\infty$ either by letting $\Omega_{\rm rot}\to\infty$ keeping $k$ fixed, or by taking $k\to 0$ at fixed $\Omega_{\rm rot}>\Omega_{\rm osc}$. We note further that by \eqref{Rm} and \eqref{scaling} the original rotational velocity $\Omega_{\rm rot}$ is related to $\Omega$ by
\beq\label{physrot}
\Omega_{\rm rot}=(sk/\gamma)^{2/(s+2)} \Omega^{(s-2)/(s+2)}.
\eeq
In particular, for the important special case $s=4$, \beq\Omega_{\rm rot}\sim \Omega^{1/3}.\eeq

The potential term and the interaction term in \eqref{gpscaled} become comparable when $\Omega\sim\eps^{-1}$. As discussed below, this is the order of the  second critical speed $\Omega_{\mathrm c2}$ above which the centrifugal force creates a hole. We are primarily interested in the case of fast rotation well above the second critical speed which means that $\Omega\gg \eps^{-1}$. 

For $\Omega\lesssim \eps^{-1}$ it is more convenient to use a different scaling than \eqref{scaling}, replacing 
$R_{\rm m}\sim \Omega_{\rm rot}^{2/(s-2)}$ by $R_\eps\sim\eps^{-2/(s+2)}$ \cite{CDY2, D}. In terms of the scaled rotational velocity $\Omega'=R_\eps^2\Omega_{\rm rot}$ the first critical velocity, where vortices start to appear, is $\Omega'_{c1}\sim |\ln\eps|$. See \cite{AD, IM1, IM2} for the case of harmonic traps and  \cite{D} for an adaption to  $s>2$. Note that $\Omega'\sim\eps^{-1}$ is equivalent to $\Omega\sim\eps^{-1}$.

\section{III. The TF density profile}

In the parameter range $\Omega\ll\eps^{-4}$ the bulk density profile of a minimizer $\psi^{\rm GP}$ of \eqref{gpscaled} can be approximately described by the Thomas-Fermi (TF) density
\beq\label{tfdens}\rho^{\rm TF}(x)=\frac{\eps^2}2\left[\mu^{\rm TF}-\gamma\Omega^2V(x)\right]_+\eeq
where $[t]_+=t$ if $t>0$ and zero otherwise. The chemical potential $\mu^{\rm TF}$ is determined by the normalization $\int \rho^{\rm TF}=1$. The density $\rho^{\rm TF}$ is the minimizer of the TF functional 
\beq \mathcal E^{\rm TF}[\rho]=\int_{\mathbb R^2}\left[\gamma\Omega^2V(x)\rho(x)+\eps^{-2}\rho(x)^2\right]\diff^2\mathbf x,\eeq i.e, the GP functional \eqref{gpscaled} without the kinetic term. The corresponding energy will be denoted $E^{\rm TF}$. For $\Omega\gtrsim \eps^{-4}$ the radial kinetic energy significantly influences the bulk density profile and the TF approximation becomes inaccurate. This will be discussed further below. 

From \eqref{tfdens} it is clear that $\rho^{\rm TF}$ vanishes at the origin for $\mu^{\rm TF}=0$ and a hole of fine radius forms as soon as $\mu^{\rm TF}<0$. The normalization of \eqref{tfdens} implies that the critical velocity  for the appearance of the hole is given by \beq\Omega_{\rm c2}=\eps^{-1}\left((2/\gamma)\hbox{$\int[-V]_+$}\right)^{-1/2}.\eeq

As $(\eps\Omega)\to\infty$ we have $\mu^{\rm TF}/(\gamma\Omega^2)\to(s-2)/2s$ and the density $\rho^{\rm TF}$ becomes concentrated around $x=1$ . The inner and outer radii, $x_{\rm in}<1$ and $x_{\rm out}>1$ respectively, of the support, as well as the chemical potential $\mu^{\rm TF}$, are determined by the equations
\beq\rho^{\rm TF}(x_{\rm in})=\rho^{\rm TF}(x_{\rm out})=0,\quad 2\pi\int_{x_{\rm in}}^{x_{\rm out}}\rho^{\rm TF}(x)\,x\,dx=1.\eeq
A Taylor expansion of $V$ around its minimum (maximum for $\rho^{\rm TF}(x)$) at $x=1$ (see Section 2.3 in \cite{CDY2} for details) gives the thickness of the support:
\beq\label{taylor}x_{\rm out}-x_{\rm in}= (\eps\Omega)^{-2/3}(12/(s-2)\gamma)^{1/3}(1+O((\eps\Omega)^{-2/3})).\eeq
By the normalization of $\rho^{\rm TF}$ it follows that the maximum $\Vert \rho^{\rm TF}\Vert_\infty=\rho^{\rm TF}(1)$ is $O((\eps\Omega)^{2/3})$. In the flat trap, on the other hand, that corresponds formally to $s=\infty$,  the thickness of the annulus where the TF density is concentrated is $O((\eps\Omega)^{-1})$ and density of  order $O(\eps\Omega)$ \cite{CY}. 

The reason for the different powers of $\eps\Omega$ can be understood by the following consideration. The Taylor expansion leading to \eqref{taylor}
is justified as soon as the turning point $x_{\rm turn}$ where $V''(x_{\rm turn})=0$ is much farther from 1 than the inner and outer radii $x_{\rm in}$ and $x_{\rm out}$, which means that
\beq\label{condition1} 1-x_{\rm turn}\gg (\eps\Omega)^{-2/3}(s-1)^{-1/3}.\eeq
Now
$x_{\rm turn}=(1/(s-1))^{1/(s-2)}$
and since we are interested in large $s$ we can write \eqref{condition1} as
$1-s^{-1/s}\gg (\eps\Omega)^{-2/3}s^{-1/3}.
$
Since
$1-s^{-1/s}=1-\exp(-(\ln s)/s)=(\ln s)/s+O((\ln s/s)^2)$
we obtain 
\beq\label{condition2} \eps\Omega\gg s/(\ln s)^{3/2}\eeq
as condition for the validity of the Taylor expansion. While this condition is always fulfilled for each finite $s$ if $\eps\Omega$ is large enough it is clearly violated for every fixed value of $\eps\Omega$ if $s\to\infty$. In fact, in  the flat trap the Taylor expansion fails and the TF density has the form
$\rho^{\rm TF}_{\rm flat}(x)\sim (\eps\Omega)^2[x^2-x_{\rm in}^2]_+$ (cf. \cite{CY}, Eq. (A.7))  with maximum value $\rho^{\rm TF}_{\rm flat}(1)\sim (\eps\Omega)$. 

In the following we consider a {\em fixed}, finite $s$ and employ formula \eqref{taylor} above. 

\section{IV. The vortex lattice regime}

An upper bound for the ground state energy $E^{\rm GP}$ of \eqref{gpscaled} can be obtained by a variational ansatz that is analogous to Eq.\ (4.1) in \cite{CY}.  It corresponds to a bulk profile determined by the TF density and a regular lattice of vortices localized at positions $\xv_i$ in the disc with radius $x_{\rm out}$ centered at the origin.  More precisely, the trial function is of the form \beq\label{latticeansatz}\psi(\xv)=c\sqrt{\rho(x)}\xi(\xv)\phi(\xv)\eeq
where $c$ is a normalization constant, $\rho$ a suitable regularization of the TF density $\rho^{\rm TF}$, $\xi(\xv)$ a 
function vanishing at the lattice points $\xv_j$ and $\phi(\xv)=\prod_j(\zeta-\zeta_j)/|\zeta-\zeta_j|$ a phase factor generated by vortices of unit strength in each lattice point. We have here used the complex notation $\zeta=x_1+\mathrm i x_2$ for points in $\mathbb R^2$. The vortices are placed so that $\nabla\phi$ compensates as far as possible the vector potential term proportional to $\Omega$ in the kinetic energy which means an arrangement in a triangular lattice with density $\Omega/\pi$. Moreover, if $t$ is the radius of a vortex core where the function $\xi$ deviates significantly from 1 the kinetic energy of a vortex localized in $\xv_j$ is, to lowest order in the small parameters,
$\sim\rho(\xv_j)|\ln(t^2\Omega)|$. Optimizing $t$ to minimize the sum of kinetic and interaction energy gives $t\sim \eps/\rho(\xv_j)^{1/2}$, provided $t$ is much smaller than the distance between vortices which is $\sim\Omega^{-1/2}$. With  $\rho(\xv_j)\sim (\eps\Omega)^{2/3}$ this  leads to $t\sim \eps^{2/3}\Omega^{-1/3}$ and this is $\ll \Omega^{-1/2}$ if $\Omega\ll\eps^{-4}$. Following closely the computation in Section 4 in \cite{CY} one now obtains for $\eps^{-1}\lesssim\Omega\ll\eps^{-4}$ the upper bound
\beq\label{upper} E^{\rm GP}\leq E^{\rm TF}+\hbox{$\frac16$}\Omega|\ln(\eps^4\Omega)|(1+O((\eps^4\Omega)^{1/3}).
\eeq
The last term is the radial kinetic energy of the density profile $\rho$. It is smaller than the second term if $\Omega\ll\eps^{-4}$.

A lower bound matching \eqref{upper} is considerably more difficult to achieve, but it can be proved using techniques from Ginzburg-Landau (GL) theory in the same way as in Section 5 in \cite{CY}. The result is

\begin{teo}[Energy between $\Omega_{\rm c2}$ and $\Omega_{\rm c3}$]
\mbox{}	\\
If $\eps^{-1}\lesssim\Omega\ll\eps^{-4}$ as $\eps\to 0$, then
\beq\label{energyas} E^{\rm GP}=E^{\rm TF}+\hbox{$\frac16$}\Omega |\ln(\eps^4\Omega)|(1+o(1)).\eeq
\end{teo}

An important difference to the flat trap considered in \cite{CY} becomes apparent here: In \cite{CY} an upper bound corresponding to \eqref{upper}  (with $\ln(\eps)$ in place of $\ln(\eps^4\Omega)$) is shown to be valid under the condition $\Omega\ll\eps^{-2}$ but the lower bound, derived by using techniques from GL theory \cite{SS},  holds only for  $\Omega\ll\eps^{-2}|\ln\eps|^{-1}$. 

For rotational speeds between the first and the second value the energy asymptotics can be proved in a similar way and is given by

\begin{teo}[Energy between $\Omega_{\rm c1}$ and $\Omega_{\rm c2}$]
\mbox{}	\\
If $|\ln\eps|\ll\Omega'\ll\eps^{-1}$ as $\eps\to 0$, then
\beq E^{ \rm GP '}\label{energyprime}= E^{ \rm TF '}+\half \Omega' |\ln(\eps^2{\Omega'})|(1+o(1)).\eeq
\end{teo}

Here $\Omega'=R_\eps^2\Omega_{\rm rot}$ as before while $E^{ \rm GP '}$  and $E^{ \rm TF '}$ denote the GP energy and the TF energy respectively, multiplied by $R_\eps^{-2}$ rather than $R_{\rm m}^{-2}$.

A further result that holds in the regimes of both Theorem 1 and Theorem 2 is that the vorticity is uniformly distributed in the bulk in the limit $\eps\to 0$.  A  proof for the case of a flat trap with Dirichlet (or Neumann) boundary conditions is given in \cite{CPRY1}, Theorem 1.3, and can be generalized to the present situation. The precise formulation of the statement in the regime of Theorem 1 is delicate because of the concentration of the density in an annulus that gets thinner as $\eps\to 0$, but the main point is that the phase circulation around a subset $\mathcal S$ with area $|\mathcal S|$ in the annulus is asymptotically equal to $2\Omega|\mathcal S|$, and this holds uniformly in $\mathcal S$ provided $|\mathcal S|$ is not too small.

\section{V. The giant vortex regime}

The first step in a study of the giant vortex transition is to consider a variational ansatz for the wave function of the form
\beq\label{ansatz} \psi(\xv)= g(\xv)\exp(\mathrm i\Omega\vartheta)\eeq
with a real valued function $g$, normalized such that $\int g^2=1$. The ansatz \eqref{ansatz} is well behaved as a function of the angular variable $\vartheta$ if $\Omega$ is an integer, otherwise it should be replaced by the integer part $[\Omega]$.  In order to simplify the notation,  however, we shall in the sequel always assume that $\Omega$ is an integer; since $\Omega\to\infty$ the inclusion of the difference $\Omega-[\Omega]$ for non-integer values leads only to negligible corrections. Inserting \eqref{ansatz} into  \eqref{gpscaled} gives
\begin{multline}\label{giantfunc}\mathcal E^{\rm GP}[\psi]=\int_{\mathbb R^2}\left\{\half|\nabla g|^2+\half\Omega^2(x-x^{-1})^2g^2\right.\\ \left. +\gamma\Omega^2\left(\hbox{$\frac 1s$}x^s-\half x^2\right)g^2+\eps^{-2}g^4\right\}\diff^2{\mathbf x}
\equiv{\mathcal E}^{\mathrm{ gv}}[{\text{\it g}}\,].
\end{multline}
The unique positive minimizer $g_{\rm gv}$ of the  functional ${\mathcal E}^{\mathrm{ gv}}[{\text{\it g}}\,]$ is  rotationally symmetric, i.e., a function of the radial variable $x$ alone. The corresponding energy will be denoted $E^{\rm gv}$. A rough upper bound for it can be obtained by taking for $g$ a regularizion of $\sqrt{\rho^{\rm TF}}$. Since $g$ is concentrated in an annulus of width $\ell=O((\eps\Omega)^{-2/3})$ and $g^2=O((\eps\Omega)^{2/3})$  the angular contribution to the kinetic energy, $\half\Omega^2\int (x-x^{-1})^2g^2$, is $O((\ell\Omega)^2)=O(\eps^{-4/3}\Omega^{2/3})$ while the radial kinetic term $\int|\nabla g|^2$ is $O((\eps\Omega)^{4/3}|\ln(\eps^4\Omega)|)$  as in \eqref{upper}.  Hence 
\beq\label{gvup} E^{\rm GP}\leq E^{\rm TF}+O(\eps^{-4/3}\Omega^{2/3})+O((\eps\Omega)^{4/3}|\ln(\eps^4\Omega)|).\eeq
From now on we shall always assume that 
\beq \label{gvregime}\Omega=\Omega_0\eps^{-4}\eeq
with some {\it fixed} $\Omega_0>0$ while $\eps\to 0$. (For the physical rotational velocity \eqref{physrot} and  $s=4$ this means that $\Omega_{\rm rot}\sim\eps^{-4/3}$.)
Then the second term in \eqref{gvup} is $O(\Omega_0^{2/3}/\eps^4)$ while the second term in \eqref{energyas} (vortex lattice kinetic energy) is
$O(\Omega_0|\ln\Omega_0|/\eps^4)$ and thus larger if  $\Omega_0$ is sufficiently large.  A radial kinetic energy term $O(\Omega_0^{4/3}|\ln\Omega_0|/\eps^4)$ is common to both \eqref{energyas} and \eqref{gvup}. The bottom line is that  for large $\Omega_0$ the giant vortex ansatz is energetically favorable to \eqref{energyas}.

These simple considerations are, however, far from a proof that a true minimizer $\psi^{\rm GP}$ of \eqref{gpscaled} has no vortices in the bulk above some $\Omega_{\rm c3}\sim \eps^{-4}$. In \cite{CRY, CPRY1} such a proof is carried out in full detail for the case of a flat trap ($s=\infty$), both with Neumann and Dirichlet boundary conditions, and it is shown that there $\Omega_{\rm c3}\sim \eps^{-2}|\ln\eps|^{-1}$. The technique used in that proof depends on tools that were originally developed in the context of  GL theory, in particular vortex ball constructions and jacobian estimates \cite{J1,JS,Sa,SS}. A prerequisite for these techniques to apply is that potential vortices can be isolated in small discs with radius much smaller than the thickness of the annulus where the bulk of the density is concentrated. As pointed out  in the discussion preceeding Eq.\ \eqref{upper} above, the radius of vortices is expected to be of order $\eps^{2/3}\Omega^{-1/3}\sim \eps^{2}\Omega_0^{-1/3}$ while the thickness of the annulus defined by the TF profile is $\sim(\eps\Omega)^{-2/3}\sim \eps^{2}\Omega_0^{-2/3}$. It is thus clear that the methods used in \cite{CRY, CPRY1} to prove the transition to a giant vortex in a `flat' trap do not apply in the present situation. Nevertheless the absence of vortices in the bulk can be proved for $\Omega$ as in \eqref{gvregime} provided $\Omega_0$ is sufficiently large.  The rest of the paper is devoted to a precise statement of this result and an outline of its  proof.

\section{VI. The gaussian density profile}

In contrast to the regime $\Omega\ll\eps^{-4}$ and the situation discussed in \cite{CRY, CPRY1} the TF profile is not a good approximation to the bulk density profile in the homogeneous trap beyond the vortex lattice regime, i.e., in the present situation for  $\Omega\sim\eps^{-4}$. In fact, the bulk of the density is contained in an annulus determined by a gaussian density distribution that we  consider first. We write the energy functional \eqref{giantfunc} as 
\begin{multline}\label{giantfunc2}{\mathcal E}^{\mathrm{ gv}}[{\text{\it g}}\,]=-\gamma\frac{(s-2)}{2s}\Omega^2\\+
\int_{\mathbb R^2}\left\{\half|\nabla g|^2+\Omega^2U(x)g^2+\eps^{-2}g^4\right\}\diff^2{\mathbf x}
\end{multline}
with
\beq U(x)=\half(x-x^{-1})^2+\gamma\left(\hbox{$\frac 1s$}x^s-\half x^2\right)+\gamma(s-2)/(2s).
\eeq
Taylor expansion of $U$ around $x=1$ (e.g. for $1/2\leq x\leq 3/2$) yields
\beq U(x)=\half\alpha^2(x-1)^2+O((x-1)^3)\eeq
with
\beq\label{alpha}
\alpha^2=4+\gamma(s-2).\eeq
We consider now for $\Omega$ as in \eqref{gvregime} 
the auxiliary one-dimensional functional
\begin{multline}\label{auxfunc}
{\mathcal E}^{\mathrm{aux}}[{\text{\it f}}\,]=\int_{\mathbb R}\left\{\half |f'|^2+\half \Omega^2\alpha^2(x-1)^2f^2+\eps^{-2}f^4\right\}{\diff} {x}\\ =\Omega\int_{\mathbb R}\left\{\half |\hat f'|^2+\half \alpha^2y^2\hat f^2+\Omega_0^{-1/2}\hat f^4\right\}{\diff} {y}
\end{multline}
where the variable transformation $y=\Omega^{1/2}(x-1)$, $\hat f(y)=\Omega^{-1/4}f(x)$ has been employed. It is clear that all three terms in \eqref{auxfunc} are of the same order of magnitude but the importance of the last term diminishes with increasing $\Omega_0$. Without the last term the minimizer is the gaussian
\beq\hat f_{\rm osc}(y)=\pi^{-1/4}\alpha^{1/4}\exp\{-\half\alpha y^2\}.\eeq

In \cite{CPRY2} it is proved that the unique positive minimizer $g_{\rm gv}$ of \eqref{giantfunc2} is well approximated by the minimizer of \eqref{auxfunc} and that the latter is, indeed, approximately gaussian for large $\Omega_0$, so that 
\beq g_{\rm gv}(x)\approx g_{\rm osc}(x)=\Omega^{1/4}\hat f_{\rm osc}(\Omega^{1/2}(x-1)).\eeq
In particular, the integral of $g_{\rm gv}^2$ over an annulus 
\beq\label{annulus}\mathcal A_{\eta}=\{\mathbf x:1-\Omega^{-1/2}\eta\leq x\leq 1+\Omega^{-1/2}\eta\}\eeq
tends to 1 if and only if $\eta\to \infty$, even though $\Omega^{-1/2}\eta\to 0$. Furthermore, the same holds for the density $|\psi^{\rm GP}|^2$ of the minimizer of the full GP functional \eqref{gpscaled}, as shown in \cite{CPRY2}. Thus any annulus of the form \eqref{annulus} contains the bulk of the density if $\eta\to\infty$. For the proof of absence of vortices in $\mathcal A_{\eta}$ it is, however,  necessary to restrict $\eta$. In fact, we prove that the annulus is vortex free if $\eta=O(|\ln\eps|^{1/2})$. In the course of the proof  slightly larger annuli, with $\eta=O(|\ln\eps|^{3/2})$ and $\eta=O(|\ln\eps|^{3/4})$ respectively, have also to be considered for technical reasons.

\section{VII. Energy estimates and absence of vortices}

Our result on the giant vortex transition is as follows:

\begin{teo}[Absence of vortices in the bulk] 
\mbox{}	\\
There are constants $0<\bar\Omega_0<\infty$ and $c>0$ such that for $\Omega=\Omega_0/\eps^4$ with $\Omega_0>\bar\Omega_0$ and $\eps$ sufficiently small the minimizer $\psi^{\rm GP}$ is free of zeros in the annulus $\mathcal A_{\rm bulk}= 
\{\mathbf x:\, |1-x|\leq c\Omega_0^{-1/2}\eps^2|\ln\eps|^{1/2}
\}.$\end{teo}
An essential part of the proof is the derivation of the precise energy asymptotics in the giant vortex regime:

\begin{teo}[Energy in the giant vortex regime] 
\mbox{}	\\
For $\Omega=\Omega_0/\eps^4$ with $\Omega_0>\bar\Omega_0$ the ground state energy is 
\begin{multline} E^{\rm GP}\label{gvenergyas}=E^{\rm gv}+O(|\ln\eps|^{9/2})=-\gamma\frac{(s-2)}{2s}\Omega^2\\+\Omega\left[\frac \alpha 2+\frac 1{2\pi}\sqrt{\frac\alpha{2\pi\Omega_0}}+O(\Omega_0^{-3/4})+O(\Omega^{-1/2})\right]\\+O(|\ln\eps|^{9/2}).
\end{multline}
\end{teo}
An upper bound to the energy is obtained by taking  in \eqref{giantfunc2}  a trial function built from the gaussian
$g_{\rm osc}(x)$. The lower bound is considerably more delicate and is discussed further below. As for the comparison with  \eqref{energyas} we note that the negative first term in \eqref{gvenergyas} is the same as the leading term in $E^{\rm TF}$, namely the potential energy in the minimum of  $\gamma\Omega^2V(x)$ at $x=1$, while the term proportional to $\Omega$ is smaller than the term $\Omega|\ln\eps^4\Omega|=\Omega\ln\Omega_0$ in \eqref{energyas} for large $\Omega_0$. 

For technical reasons we consider besides the functional \eqref{giantfunc2} also a functional 
$\mathcal E^{\rm gv}_\eta$ defined by the same formula except that the integration is restricted to an annulus $\mathcal A_\eta$ with $\eta=O(|\ln\eps|^{3/2})$. Its unique positive minimizer $g_\eta$ can be shown to be close to $g_{\rm osc}$ on the smaller annulus $\mathcal A_{\sqrt{\eta}}$\ :
\beq\label{gaussianapprox} g_\eta(x)=(1+O(\Omega_0^{-1/4}))g_{\rm osc}(x).\eeq
The corresponding energy $\mathcal E^{\rm gv}_\eta(g_\eta)$ is denoted by $E^{\rm gv}_\eta$. The choice of $\eta=O(|\ln\eps|^{3/2})$ is to some extent arbitrary but the method of sub- und supersolutions \cite{E} used in the proof of \eqref{gaussianapprox} and of the exponential smallness of $|\psi^{\rm GP}|$ outside of  $\mathcal A_{\sqrt{\eta}}$ (that is needed for the energy estimates) requires that $|\ln\eps|\ll\eta\ll\eps^{-1}$.

The next step is a decoupling of the energy functional that has been used repeatedly in analogous contexts in GL and GP theory \cite{LM}. We define for $\mathbf x \in\mathcal A_\eta$ a function $u(\mathbf x)$ by writing
\beq\psi^{\rm GP}\label{factorization}(\mathbf x)=g_\eta(x)u(\mathbf x)\exp(\mathrm i\Omega\vartheta).\eeq
Since $g_\eta$ is without zeros, the function $u$ contains all possible zeros of the minimizer $\psi^{\rm GP}$ in the annulus. The variational equation for $g_\eta$ 
leads to the lower bound
\beq E^{\rm GP}\geq E^{\rm gv}_\eta+\mathcal E_\eta[u]\eeq
with
\beq\label{eeta} \mathcal E_\eta[u]=\int_{\mathcal A_\eta}g_\eta^2\left\{\half|\nabla u|^2-\mathbf B\cdot \mathbf J(u)+\eps^{-2}g_\eta^2(1-|u|^2)^2\right\}\eeq
where $\mathbf B=\Omega(x-x^{-1})\mathbf e_\vartheta$ and $\mathbf J(u)=\frac {\rm i}2(u\nabla u^*-u^*\nabla u)$.

The main task is now to estimate the negative term involving $g_\eta^2 \mathbf B\cdot \mathbf J(u)$. As usual in the context of GP theory (see, e.g., \cite{AAB,CRY, CPRY1}) an essential step is an integration by parts. Namely, one writes
$g_\eta^2 \mathbf B=\nabla^\perp F$ with the dual gradient $\nabla^\perp=(-\partial_{x_2},\partial_{x_1})$ and a potential function $F$. In order to employ \eqref{gaussianapprox} we also restrict the integration to $\mathcal A_{\sqrt \eta}$ that can be shown to create only negligible errors if $\eta\gg|\ln\eps|$. If $g_\eta|\mathbf B|$ would be exactly symmetrical about $x=1$ like $g_{\rm osc}$ we could choose $F$ to vanish on both boundaries of the annulus $\mathcal A_{\sqrt \eta}$ and integration by parts would give
\beq\label{intj} -\int_{\mathcal A_{\sqrt{\eta}}} g_\eta^2\, \mathbf B\cdot \mathbf J(u)=\int_{\mathcal A_{\sqrt{\eta}}} F\,\nabla^\perp\cdot\mathbf J(u).\eeq
Moreover, a simple computation employing \eqref{gaussianapprox} gives $|F(x)|\leq \alpha^{-1}(1+O(\Omega_0^{-1/4})g_\eta^2(x)$, while $|\nabla^\perp\cdot\mathbf J(u)|\leq |\nabla u|^2$. Thus, because $\alpha>2$, the positive first term in \eqref{eeta} integrated over 
${\mathcal A_{\sqrt{\eta}}}$ dominates \eqref{intj}  for $\Omega_0$ large enough.

This reasoning is, however, not rigorous because $g_{\eta}|\mathbf B|$ is not perfectly symmetric about $x=1$ and if $F$ is chosen to vanish on one boundary, e.g., the inner one, it will not vanish exactly on the other. The integration by parts then creates a boundary term 
$ F(R) \oint_{x=R} \mathbf J(u)\cdot \diff{\mathbf\ell}$
with $R$ the radius of that boundary.   To control the circulation integral one would like to use a part of positive kinetic energy $\int_{\mathcal A_{\sqrt\eta}} g_\eta^2|\nabla u|^2$ so the first step is to transform the boundary integral into  two-dimensional integrals.
If $\bar R=R-c\Omega^{-1/2}$ with $c$ small and $\chi$ is a smooth, monotone  radial function on $[\bar R,R]$ with $\chi(\bar R)=0$,  $\chi(R)=1$ and $|\nabla^\perp \chi|\leq C\Omega^{1/2}$, we can write
\beq\label{jsplit} \oint_{x=R} \mathbf J(u)\cdot {\diff}{\mathbf\ell}=\int \nabla^\perp\chi\cdot \mathbf J(u)+\int \chi\,\nabla^\perp\cdot \mathbf J(u).\eeq
The analogous computation holds for $\bar R=R+c\Omega^{-1/2}$ and the interval $[\bar R,R]$. To bound the integrals in terms of $\int g_\eta^2|\nabla u|^2$, however, we need $g_\eta$ to be large in the interval $[\bar R,R]$ (or $[ R,\bar R]$). Because $g_\eta$ is, in fact,  very small on the boundary of the annulus $\mathcal A_{\sqrt\eta}$  this strategy runs into difficulties. 

A way out is to introduce {\em two} potential functions, $F_1$ vanishing on the inner boundary of the annulus  and $F_2$, vanishing on the outer boundary. The former is applied below the radius $R_{\rm max}$ where $g_\eta$ has its maximum, the latter above $R_{\rm max}$. The integration by parts now creates two boundary terms of opposite signs at $x=R_{\rm max}$, and  one has to estimate 
\beq \left[F_1(R_{\rm max})-F_2(R_{\rm max})\right]\int_{x=R_{\rm max}} \mathbf J(u)\cdot \diff{\mathbf\ell}.\eeq
Since $g_\eta$ is large in a neigbourhood of $R_{\max}$ Eq.\ \eqref{jsplit} can be put to good use: Taking $R=R_{\rm max}$ we have $g_\eta^2\geq C\Omega^{1/2}$ on $[\bar R,R]$ by \eqref{gaussianapprox}, 
and from \eqref{jsplit}, using the normalization of $g^2|u|^2$, we obtain
\begin{multline}\label{circest}
\Big| \oint_{x=R_{\rm max}} \mathbf J(u)\cdot {\diff}{\mathbf\ell}\Big|\leq C\int_{\mathcal A_{\sqrt\eta}}g_\eta^2|u||\nabla u|\\
+C\Omega^{-1/2}\int_{\mathcal A_{\sqrt\eta}}g_\eta^2|\nabla u|^2\\\leq C(\Omega^{-1/2}+\delta)\int_{\mathcal A_{\sqrt\eta}}g_\eta^2|\nabla u|^2+C\delta^{-1}
\end{multline}
for any $\delta>0$. Here and in the following $C$ denotes a finite, positive  constant that may differ from line to line. The difference $\big |F_1(R_{\rm max})-F_2(R_{\rm max})\big |$ is estimated separately. It is small because $g_\eta|\mathbf B|$ is approximately symmetric  about $x=1$, and making use of the variational equation for $g_\eta$ it is shown to be at most $O(\Omega_0\eta^{3/2})$. Choosing $\delta=C_\delta\Omega_0^{-1}\eta^{-3/2}$ with sufficiently small $C_\delta$ we now obtain for all  $\Omega_0>\bar\Omega_0$ sufficiently large the crucial bound
\begin{multline}\label{crucial} \int_{\mathcal A_{\sqrt\eta}}g_\eta^2\left\{\half|\nabla u|^2-\mathbf B\cdot \mathbf J(u)\right\} \geq
 -C\Omega_0^2\eta^3\\+C'\int_{\mathcal A_{\sqrt\eta}}g_\eta^2|\nabla u|^2.
\end{multline}
with $C'>0$. Replacing the integration domain by $\mathcal A_{\eta}$ only produces a negligible correction if $\eta\gg |\ln\eps|$, and $E^{\rm gv}_\eta$ also differs from $E^{\rm gv}$ only by small terms. This completes the proof of the lower bound in Theorem 4. 

A further consequence of \eqref{crucial}, combined with the variational bound $E_\eta^{\rm gv}\leq 0$ and the exponential smallness of $|\psi^{\rm GP}|^2$ outside $\mathcal A_{\sqrt{\eta}}$, is the bound 
\beq \label{intbound}\int_{\mathcal A_{\sqrt\eta}}\eps^{-2}g_\eta^4(1-|u|^2)^2\leq C\Omega_0^2\eta^3\eeq
on the interaction term for $\Omega_0>\bar \Omega_0$. This leads to Theorem 3 by the following reasoning. Using the variational equation satisfied by $g_\eta$ as well as the Gagliardo-Nirenberg inequality in a similar way as in \cite{CRY}, Lemma 5.1., one obtains the gradient estimate \beq\label{gradest} |\nabla u(\mathbf x)|\leq C\eps^{-2+(c^2\alpha/2)}\eeq
with $\alpha$ as in \eqref{alpha} for all $\mathbf x$ such that \beq \label{xcond} |1-x|\leq c\Omega_0^{-1/2}\eps^2|\ln\eps|^{1/2}.\eeq Here it has been used that as a consequence of  \eqref{gaussianapprox} we have \beq g_\eta(x)\geq C\eps^{-1+(c^2\alpha/2)}\eeq
for $x$ satisfying \eqref{xcond}.

We now claim that  as $\eps\to 0$, $|1-|u(\mathbf x)||< |\ln \eps|^{-a}$ holds for all $a>0$ on the annulus defined by \eqref{xcond}, provided $c<(2/\alpha)^{1/2}$. The proof is by contradiction: 
Suppose  that $|1-|u(\mathbf x)||\geq |\ln \eps|^{-a}$ at some $\mathbf x$ and $a>0$. Then the gradient estimate \eqref{gradest} implies that $|1-|u(\mathbf x)||\geq |\ln \eps|^{-a}/2$ on a disk of radius $C\eps^{2-(c^2\alpha/2)}$ around $\mathbf x$.  We thus obtain
\beq  \int_{\mathcal A_{\sqrt\eta}}\eps^{-2}g_\eta^4(1-|u|^2)^2\geq C\eps^{-2+c^2\alpha} |\ln\eps|^{-2a}.\eeq
This is a contradiction to \eqref{intbound} for $c<(2/\alpha)^{1/2}$ and $\eta=|\ln\eps|^{3/2}$. Thus $|1-|u(\mathbf x)||< |\ln \eps|^{-1}$ holds, implying that $u$, and hence also $\psi^{\rm GP}$, is free of zeros in the bulk defined by \eqref{xcond}. 

\section{VIII. Circulation and symmetry breaking}

The degree (winding number) of the giant vortex ansatz \eqref{ansatz} is clearly $\Omega$. This can also be shown to hold, to very good accuracy, for the true minimizer  $\psi^{\rm GP}$ in the giant vortex regime, ensuring a macroscopic circulation around the central hole where the density is strongly depleted: 
\begin{teo}[\textbf{Asymptotics for the degree}]
		\label{teo: gv degree}	
		\mbox{}	\\
		If $ \Omega $ is given by \eqref{gvregime} with $ \Omega_0 > \bar\Omega_0 $ and $R$ is any radius satisfying  $
 R = 1 + O(\Omega^{-1/2})$
then as $ \eps \to 0 $ the degree of $\psi^{\rm GP}$ around the circle with radius $R$ is  $\Omega  + O(\Omega_0 |\ln \eps| ^{9/4})$
		\end{teo}
		Indeed, a simple computation, using \eqref{factorization} gives
		\beq\text{degree of }\psi^{\rm GP}=\Omega+{\rm i}(2\pi)^{-1}\oint_{x=R}u^{-1}|u|\partial_{R\vartheta}(u|u|^{-1})
		\eeq
and the second term is easily estimated exploiting Eq.\  \eqref{circest}  and the bound $\int g_\eta^2|\nabla u|^2\leq C\Omega_0^2\eta^3$ that follows from \eqref{crucial} together with $E^{\rm gv}_\eta\leq 0$.	

According to the previous Theorems 3-5 the ansatz \eqref{ansatz} gives an excellent approximation to the energy and the qualitative properties of a true minimizer $\psi^{\rm GP}$ if $\Omega_0$ is large enough. Nevertheless, while \eqref{ansatz} is an eigenfunction of angular momentum and its modulus therefore rotationally symmetric, this is not the case for a true minimizer:	
		\begin{teo}[\textbf{Symmetry breaking}]
		\label{teo: symmetry breaking}	
		\mbox{}	\\
No minimizer $\psi^{\rm GP}$ in the giant vortex regime $\Omega_0>\bar \Omega_0$ is an eigenfunction of angular momentum.
\end{teo}
The indirect proof is very similar to the proof of a corresponding result in \cite{CPRY1}, Theorem 1.6, that in turn is inspired by Theorem 2 in \cite{Sei}, and \cite{M}. One assumes that $\psi^{\rm GP}(\xv)=f(x)\exp({\mathrm i n \vartheta})$ with a real radial function $f$ and $n\in\mathbb Z$. As a byproduct of the analysis in the previous section it can be shown that $n=\Omega (1+O(\eps^4|\ln\eps|^{9/4})$. Moreover, $f$ is exponentially small w.r.t.\ $\eps$ outside the annulus $\mathcal A_{\sqrt \eta}$ with $\eta=|\ln\eps|^{3/2}$. It has a unique maximum at $x_{\rm m}$, close to 1. Define
$$A(x)=x^2f'(x) \text{ for } \xv\in \mathcal A_{\sqrt \eta},\, x\leq x_{\rm m}$$
and $A(x)=0$ for $x>x_{\rm m}$ as well as a smooth interpolation to zero inside the inner boundary  of $\mathcal A_{\sqrt\eta}$. Likewise define 
$$B(x)=nxf(x) \text{ for } \xv\in \mathcal A_{\sqrt \eta}$$ with a smooth interpolation to 0 outside $\mathcal A_{\sqrt\eta}$.
Consider then 
\begin{multline}w(\xv)=(A(x)+B(x))\exp(\mathrm i(n+1)\vartheta)\\+(A(x)-B(x))\exp(\mathrm i(n-1)\vartheta).\end{multline}
A computation, employing the variational equation for $f$,  then shows that the second variation of the GP functional, i.e., the quadratic form $Q$ given by Eq. (2.3) in \cite{Sei}, is negative when evaluated on $w$, implying that $f(x)\exp({\mathrm i n \vartheta})$ can not be a minimizer.

\section{IX. Conclusions}

We have analyzed the change in the density and vortex patterns of a superfluid in a rotating, anharmonic trap as the rotational velocity and the interaction parameter both tend to infinity. In particular we have shown rigorously that the fluid undergoes a transition into a giant vortex state where there are no vortices in the bulk if the rotational velocity exceeds a certain limit depending on the interaction strength while a macroscopic circulation remains. In the paper we have focused on \lq soft' trapping potentials, e.g., the sum of a quartic and a quadratic potential, where the problem turns out to differ markedly, both physically and mathematically,  from the previously considered case of a flat trapping potential with a fixed boundary. The differences concern both the shape of the bulk density in the giant vortex state, that in soft traps turns out to be approximately gaussian rather than of  Thomas-Fermi type, as well as the relative size of vortex cores and the annulus where the bulk is concentrated. \bigskip

{\small {\bf Acknowledgements.} MC and NR acknowledge the hospitality of the Erwin Schr\"odinger Institute for Mathematical Physics in Vienna where part of this work was carried out. The work of NR was supported by the European Research Council under the European Community Seventh Framework Programme (FP7/2007-2013 Grant Agreement MNIQS no. 258023), the work of MC by the same programme under the Grant Agreement CoMBoS no. 239694.}
%%%%%%%%%%%%%%%%%%%%%%%%%%%%%%%%%%%%%


\begin{thebibliography}{99}
\bibitem[1]{IM1} 	\textsc{R. Ignat, V. Millot}, \emph{J. Funct. Anal.} \textbf{233} (2006), 260--306.


\bibitem[2]{AJR} \textsc{A. Aftalion, R.L. Jerrard, J. Royo-Letelier},  \emph{J. Func. Anal.} \textbf{260} (2011), 2387--2460. 

\bibitem[3]{A} 		\textsc{A. Aftalion}, {\it Vortices in Bose-Einstein Condensates}, Progress in Nonlinear Differential Equations and their Applications \textbf{67}, Birkh\"auser, Basel, 2006.

\bibitem[4]{Fe1} 	\textsc{A.L. Fetter},  \emph{Rev. Mod. Phys.} \textbf{81} (2009), 647--691.




\bibitem[5]{Co} 		\textsc{N.R. Cooper},  \emph{Adv. Phys.} \textbf{57} (2008), 539--616. 



\bibitem[6]{CD}  \textsc{Y. Castin, R. Dum},
{\emph Eur. Phys. J.
D} {\bf 7} (1999), 399.


\bibitem[7]{AD}  	\textsc{A. Aftalion, Q. Du},  {\it Phys. Rev. A} \textbf{64} (2001), 063603a.

\bibitem[8]{IM2} 	\textsc{R. Ignat, V. Millot},  \emph{Rev. Math. Phys.} \textbf{18} (2006), 119--162.



\bibitem[9]{FB} 		\textsc{U.R. Fischer, G. Baym},  \emph{Phys. Rev. Lett.} \textbf{90} (2003), 140402.




\bibitem[10]{CDY1}   	\textsc{M. Correggi, T. Rindler-Daller, J. Yngvason},  \emph{J. Math. Phys.} \textbf{48} (2007), 042104.

\bibitem[11]{CY}		\textsc{M. Correggi, J. Yngvason}, {\it J. Phys. A: Math. Theor.} {\bf 41} (2008), 445002.

\bibitem[12]{Fe2}	\textsc{A.L. Fetter},   \emph{Phy. Rev. A} \textbf{64} (2001), 063608. 


\bibitem[13]{FJS} 	\textsc{A.L. Fetter, B. Jackson, S. Stringari},  \emph{Phys. Rev. A} \textbf{71} (2005), 013605. 

\bibitem[14]{FZ} 		\textsc{H. Fu, E. Zaremba},   \emph{Phys. Rev. A} \textbf{73} (2006), 013614. 

\bibitem[15]{KTU} 	\textsc{K. Kasamatsu, M. Tsubota, M. Ueda},   \emph{Phys. Rev. A} \textbf{66} (2002), 053606.

\bibitem[16]{KF} 		\textsc{J.K. Kim, A.L. Fetter},   \emph{Phys. Rev. A} \textbf{72} (2005), 023619.



\bibitem[17]{DK}  \textsc{I. Danaila, P. Kazemi},   \emph{SIAM J. Sci. Comp.} \textbf{32} (2010), 2447.

\bibitem[18]{KB} 		\textsc{G.M. Kavoulakis, G. Baym},  \emph{New J. Phys.} \textbf{5} (2003), 51.1--51.11.

\bibitem[19]{R1} 		\textsc{N. Rougerie},  \emph{J. Math. Pures Appl.} \textbf{95} (2011), 296--347.



\bibitem[20]{CRY}	\textsc{M. Correggi,  N. Rougerie, J. Yngvason}, \emph{Commun. Math. Phys.} \textbf{303} (2011), 451--508.



\bibitem[21]{CPRY1}	\textsc{M. Correggi, F. Pinsker, N. Rougerie, J. Yng\-va\-son},  \emph{J. Stat. Phys.}, \textbf{143} (2011), 261--305.

\bibitem[22]{CPRY2}	\textsc{M. Correggi, F. Pinsker, N. Rougerie, J. Yngvason}, Critical Rotational Speeds for Superfluids in Homogeneous Traps, \emph{arXiv:\/} 1108.5429.


\bibitem[23]{R2}		\textsc{N. Rougerie}, \emph {Arch. Rational Mech. Anal.}, published online (2011).



\bibitem[24]{BSSD} \textsc{V. Bretin, S. Stock, Y. Seurin, J. Dalibard}, {\it Phys. Rev. Lett.} {\bf 92} (2004), 050403. 




\bibitem[25]{Ry}           \textsc{C. Ryu, M.F. Andersen, P. Clade, V. Natarajan, K. Helmerson, W.D. Phillips},  \emph{Phys. Rev. Lett.} \textbf {99}  (2007), 260401.



\bibitem[26]{H}   \textsc{K. Henderson, C. Ryu, C. Mac Cormick, M.G. Boshier},  \emph{New J. Phys.} \textbf{11} (2009), 043030. 


 


\bibitem[27]{Sh}\textsc{B.E. Sherlock, M. Gildemeister, E. Owen, E. Nugent, C.J. Foot}, \emph{Phys. Rev. A} \textbf{83} (2011), 043408.






\bibitem[28]{AAB}	\textsc{A. Aftalion, S. Alama, L. Bronsard},  {\it Arch. Rational Mech. Anal.} {\bf 178} (2005), 247--286.

\bibitem[29]{LS} 		\textsc{E.H. Lieb, R. Seiringer},  \emph{Comm. Math. Phys.} \textbf{264} (2006), 505--537.



\bibitem[30]{CDY2}   	\textsc{M. Correggi, T. Rindler-Daller, J. Yngvason},  \emph{J. Math. Phys.} \textbf{48} (2007), 102103.

\bibitem[31]{D}\textsc{T. Rindler-Daller}, \emph{Physica} A, \textbf{387} (2008), 1851--1874.




\bibitem[32]{J1} 		\textsc{R.L. Jerrard},  \emph{SIAM J. Math. Anal.} \textbf{30} (1999), 721--746.

\bibitem[33]{Sa} 		\textsc{E. Sandier},  \emph{J. Funct. Anal.} \textbf{152} (1998), 349--358.


\bibitem[34]{JS} 		\textsc{R.L. Jerrard, H.M. Soner},  \emph{Calc. Var. Partial Differential Equations} \textbf{14} (2002), 524--561.

\bibitem[35]{SS} 		\textsc{E. Sandier, S. Serfaty}, \emph{Vortices in the Magnetic Ginzburg-Landau Model}, in Progress in Nonlinear Differential Equations and their Applications \textbf{70}, Birkh\"{a}user, Basel, 2007.



\bibitem[36]{E}	\textsc{L.C. Evans}, {\it Partial Differential Equation}, Graduate Studies in Mathematics {\bf 19}, AMS, Providence, 1998.

\bibitem[37]{LM} 		\textsc{L. Lassoued, P. Mironescu},  \emph{J. Anal. Math.} \textbf{77} (1999), 1--26.


\bibitem[38]{Sei} \textsc{R. Seiringer}, \emph{Commun. Math. Phys.} \textbf{229} (2002), 491--509.


\bibitem[39]{M} \textsc{P. Mironescu},  \emph{J. Funct. Anal.} \textbf{130}  (1995), 334--344.



\bibitem[40]{BB1} \textsc{I. Bialynicki-Birula}, \emph{Phys. Rev. A}
{\bf 61} (2000), 032110 

\bibitem[41]{BB2} \textsc{I. Bialynicki-Birula}, \emph{Phys. Rev. A}
{\bf 65} (2002), 014101 


\bibitem[42]{BBS} \textsc{I.\ Bialynicki-Birula, T. Sowinski}, in:  {\it Nonlinear Waves: Classical and Quantum Aspects}, F. Kh. Abdullaev and
V. V. Konotop (eds.), Kluver, Amsterdam, 2004, p. 99























\end{thebibliography}
\end{document}